\documentstyle[12pt, psfig, twoside]{article}
\def\references{\section*{References}
\bgroup\parindent=0pt\parskip=.5ex
\def\refpar{\par\hangindent=2em\hangafter=1}}
\def\endreferences{\refpar\egroup}
\def\reference{\relax\refpar}

\newcommand\nab {   \mbox{\boldmath{$\nabla$}}  }
\newcommand\tim {   \mbox{\boldmath{$\times$}}  }

\title{Theory of Turbulent Accretion Disks}
\author{Caroline E. J. M. L. J. Terquem\\ \\ 
{\normalsize {\em Institut d'Astrophysique de Paris}} \\ 
{\normalsize {\em and Universit\'e Denis Diderot--Paris ~VII}} \\ \\ 
{\small{\em To appear in the proceedings of the Aussois 2000
summer school}} \\ {\small{\em 
``Formation Stellaire et Physique des Etoiles Jeunes''}}}

\date{}
\begin{document}
\maketitle

\begin{abstract}

In low--mass disks, turbulent torques are probably the most important
way of redistributing angular momentum.  Here we present the theory of
turbulent accretion disks.  We show the molecular viscosity is far too
small to account for the evolutionary timescale of disks, and we
describe how turbulence may result in enhanced transport of (angular)
momentum.  We then turn to the magnetorotational instability, which
thus far is the only mechanism that has been shown to initiate and
sustain turbulence in disks.  Finally, we present both the basis and
the structure of $\alpha$ disk models.

\end{abstract}

\section{Introduction}

It is an observational fact that at least part of the mass in
protostellar disks is accreted onto the central objet.  In a Keplerian
accretion disk, the specific angular momentum increases with radius.
Therefore, a particle can be accreted only if its angular momentum is
removed or transferred to particles located at lager radii.  Whether
angular momentum is removed from or redistributed within the disk
depends on whether the disk is subject to external or internal
torques.  Possible external torques can either be magnetic (when an
outflow is present) or tidal (in binary systems), whereas possible
internal torques can either be gravitational (massive disks) or
turbulent.  These mechanisms have been discussed by Papaloizou \& Lin
(1995).

During the early stages of disk evolution, when the disk is still
embedded (class~0/1 object) and has a significant mass compared to the
central star, there may exist strong disk winds and bipolar outflows
(e.g. Reipurth et al. 1997) with associated magnetic fields.  During
this stage, a hydromagnetic disk wind may be an important means of
angular momentum removal for the system (Blandford \& Payne 1982).
Because of the action of magnetic torques, material ejected from the
disk is able to carry away significantly more angular momentum than it
originated with in the disk.  Therefore, even a modest ejection rate
can lead to a significant accretion rate through the disk.  However,
observations indicate that outflows may exist only in the early stages
of disk evolution, so that this mechanism cannot account for angular
momentum transport during the whole life of accretion disks.  In
addition, it may affect only the inner parts.

When the mass of the disk is significant compared to that of the star,
gravitational instabilities may develop, leading to outward angular
momentum transport (Papaloizou \& Savonije 1991; Heemskerk et
al. 1992; Laughlin \& Bodenheimer 1994; Pickett et al. 1998) that
results in additional mass growth of the central star.  This
redistribution of mass may occur on the dynamical timescale (a few
orbits) of the outer part of the disk and thus may be quite rapid: on
the order of $10^5$~yr for a disk radius of 500~AU.  The parameter
governing the importance of disk self--gravity is the Toomre
parameter, $Q \sim M_{\ast} H/(M_d r)$, with $M_{\ast}$ being the
central mass, $M_d$ being the disk mass contained within radius $r$
and $H$ being the disk semi--thickness.  Typically $H/r \sim 0.1$
(Stapelfeldt et al. 1998) such that the condition for the importance
of self--gravity, $Q \sim 1$, gives $M_d \sim 0.1 M_{\ast}$.  During
the period of global gravitational instability, it is reasonable to
suppose that the disk mass is quickly redistributed and reduced by
accretion onto the central object such that the effects of
self--gravity become negligible.

If the disk surrounds a star which is in a pre--main sequence binary
system, tidal torques transport angular momentum outward, from the
disk rotation to the orbital motion of the binary.  However, although
tidal effects are important for truncating protostellar disks and for
determining their size, it is unlikely that tidally--induced angular
momentum transport plays a dominant role in the evolution of
protostellar disks (see Terquem~2001 and references therein).  In a
non self--gravitating disk, the amount of transport provided by tidal
waves is probably too small to account for the lifetime of
protostellar disks.  In addition, tidal effects tend to be localized
in the disk outer regions.

When the disk mass is such that self--gravity can be ignored and the
jet activity has significantly decreased, turbulent torques may become
the most important way of redistributing angular momentum in the disk.
Historically, the first angular momentum transport mechanism to be
considered was through the action of viscosity (von Weizs\"acker
1948).  However, in order to result in evolution on astronomically
interesting timescales, it is necessary to suppose that an anomalously
large viscosity is produced through the action of some sort of
turbulence.

In this paper we present the theory of turbulent accretion disks.  We
first calculate the molecular viscosity in a disk, and show this is
far too small to account for the evolutionary timescale of disks.  We
then describe how turbulence may result in enhanced transport of
(angular) momentum.  About ten years ago, a linear magnetic
instability was identified (Balbus \& Hawley 1991) that results in
MHD turbulence.  We present this instability.  We then describe both
the basis and the structure of $\alpha$ disk models, which were
introduced by Shakura \& Sunyaev (1973).

\section{Fundamental of hydrodynamics}

\subsection{Fluid mechanics vs. kinetic theory}

The fundamental equations of fluids can be derived by considering them
as either a collection of particles (kinetic theory) or as a smooth
continuum.  This latter approach is justified when the mean free path
$\lambda$ of the particles is very small compared to the macroscopic
length scale $L$ of interest in the fluid.  This condition is not met
in stellar systems (galaxies or star clusters), where $\lambda \gg L$,
nor in planetary rings, where $\lambda \sim L$, but it is fulfilled in
disks.  The conservation equations are obtained more straightforwardly
when considering the fluid as a continuum, which is the approach we
will use here.

This does not lead to a precise expression for the transport
coefficients, however, in contrast to kinetic theory.  Transport of
energy, mass and momentum occurs in a gas which is out of equilibrium
(i.e. in which the distribution function is not a Maxwell--Boltzmann
distribution) through molecular collisions. Most of the time, the
departure from equilibrium is tiny, so that the distribution function
is {\em nearly} maxwellian.  Within the context of the kinetic theory,
in which molecular collisions are explicitly calculated, the
Chapman--Enskog procedure gives the transport coefficients by
considering small variations of the distribution function around the
Maxwell--Boltzmann distribution.
Such a calculation is not possible when fluids are viewed as continua,
as in this case molecular collisions are not explicitly calculated.
It is possible however to get a phenomenological expression for the
transport coefficients in this context, as we shall see below.

For more details on the kinetic theory of gases, the reader is
referred to, e.g., Huang (1987) and Shu (1992).  For the description
of fluids as continua, which is developed below, a more complete
presentation can be found, e.g., in Acheson (1990), Batchelor (1967),
Landau \& Lifchitz (1987) and Shu (1992).

\subsection{Conservation laws}

\subsubsection{Mass conservation}

If there is no sink or source of matter, the time rate of mass
variation within a fixed volume $V$ is equal to minus the flux of mass
(mass advected by the flow) through the surface $S$ which encloses the
volume:

\begin{equation}
\frac{\partial}{\partial t} \int_V \rho dV = - \int_S \rho {\bf v}
\cdot {\bf n} dS,
\end{equation}

\noindent where $\rho({\bf r},t)$ and ${\bf v}({\bf r},t)$ denote the
density and velocity vector of the fluid at location ${\bf r}$ and
time $t$, and ${\bf n}$ is the unit normal to the surface oriented
outward.

Since this has to be satisfied for any volume $V$, and using the
divergence theorem, we get:

\begin{equation}
\frac{\partial \rho}{\partial t} + \nab \cdot \left( \rho {\bf v}
\right) = 0.
\label{mass1}
\end{equation}

The mass conservation equation derived above is also know as the {\em
continuity} equation.  It states that mass variation in a fixed volume
is due to a continuous flow of mass through the surfaces which enclose
the volume, and not to 'jumps' from one location to another remote
one.

Equation~(\ref{mass1}) has been derived by considering the variation
of mass of a volume fixed in space.  There are two contributions to
the time rate of mass change in this volume: $\rho \nab \cdot {\bf
v}$, which is due to the change of the volume itself, and ${\bf v}
\cdot \nab \rho$, which is due to the mass flowing through the volume.

We can also consider the time rate of mass change of a volume of fluid
that we follow along its trajectory.  For that we define the Lagrangian
derivative $d/dt$, which is the time rate of change in a frame moving
with a particular fluid element along its trajectory. The density
$\rho$ depends both on ${\bf r}$ and $t$ and, if ${\bf r}$ is the
location of the moving fluid element, it depends also on $t$. Then

\begin{equation}
\frac{d \rho}{dt} = \frac{\partial \rho}{\partial t} + \frac{\partial
\rho}{\partial x_i} \frac{d x_i}{dt} = \frac{\partial \rho}{\partial
t} + {\bf v} \cdot \nab \rho ,
\label{lagrange}
\end{equation}

\noindent were the $x_i$ ($i=1,2,3$) denote the coordinates and we
adopt the standard Einstein notation of summation over repeated
indices.

Equation~(\ref{mass1}) can therefore be rewritten as:

\begin{equation}
\frac{d \rho}{dt} + \rho \nab \cdot {\bf v} = 0.
\label{mass2}
\end{equation}

\noindent This expresses the fact that the mass change in a volume
moving with the fluid is due only to the change in the volume.

\subsubsection{Incompressibility}

A fluid is said to be incompressible if a volume element is neither
compressed nor dilated when moving with the flow.  Let us consider a
fixed volume $V$ of the fluid bounded by the surface $S$.  The volume
of fluid which leaves $V$ through an infinitesimal surface $dS$ per
unit time is ${\bf v} \cdot {\bf n} dS$, where ${\bf n}$ is the unit
normal to the surface oriented outward.  Therefore

\begin{equation}
\frac{dV}{dt} = \int_S {\bf v} \cdot {\bf n} dS = \int_V \left( \nab
\cdot {\bf v} \right) dV,
\end{equation}

\noindent where we have used the divergence theorem to get the final
expression.  The condition for the fluid to be incompressible is therefore:

\begin{equation}
\nab \cdot {\bf v} = 0,
\end{equation}

\noindent or, equivalently, using the mass conservation
equation~(\ref{mass2}):

\begin{equation}
\frac{d \rho}{dt} = 0.
\end{equation}

This expresses the fact that, in an incompressible fluid, the density
of a fluid element is constant along its trajectory.

\subsubsection{Equation of motion and momentum conservation}

The product of the acceleration of a fluid element moving with the
flow, $d{\bf v}/dt$, by the mass density $\rho$ is equal to the net
force per unit volume exerted onto this element.  This has
contribution from external forces (e.g., gravitational, magnetic...)
and/or from momentum transport within the flow due to molecular
collisions (pressure and viscous forces).  The local form of the
equation of motion is then:

\begin{equation}
\rho \frac{d{\bf v}}{dt} = \rho \left( \frac{\partial {\bf v}}{
\partial t} + \left( {\bf v} \cdot \nab \right) {\bf v} \right)= -
\nab P + {\bf F}_{visc} + {\bf F} ,
\label{motion}
\end{equation}

\noindent where $P$ is the pressure, ${\bf F}_{visc}$ is the viscous
force per unit volume and ${\bf F}$ is any other force per unit volume
which does not arise from pressure and viscosity.  When ${\bf
F}_{visc} = {\bf 0}$, this equation is known as {\em Euler
equation}.

Using equation~(\ref{mass1}), we can rewrite the equation of motion
under the form:

\begin{equation}
\frac{\partial}{\partial t} \left( \rho {\bf v} \right) = - \nab \cdot
\left( \rho {\bf v} {\bf v} \right) - \nab P + {\bf F}_{visc} + {\bf
F},
\label{momentum}
\end{equation}

\noindent where, according to our notation, the $i$ component of the
vector $\nab \cdot \left( \rho {\bf v} {\bf v} \right)$ is $\nab \cdot
\left( \rho v_i {\bf v} \right)$.

This is the momentum conservation equation, which states that the time
rate of momentum variation within a fixed volume is equal to minus the
momentum which is advected by the flow through the surface which
encloses the volume plus the net force exerted onto the volume.

\subsection{Viscous forces: Molecular transport of momentum}
\label{sec:viscousforces}

The viscous force in a fluid is due to an irreversible transport of
momentum from regions where the velocity is higher to regions where it
is lower.  A variational calculation shows that uniform rotation is an
extremum for the energy.  Sometimes it is a minimum, in which case
dissipation of energy at fixed (angular) momentum results in uniform
rotation (e.g.  synchronous binaries).  This tends to be the case when
pressure is dominant against gravity.  Sometimes the extremum is a
maximum, in which case all the matter tends to accumulate at the
center, with all the angular momentum at infinity. This tends to be
the case when rotation is dominant against gravity, and this is what
happens in disks (for further details, see Lynden--Bell \& Kalnajs
1972; Lynden--Bell \& Pringle 1974).  The dissipation of energy
results from molecular collisions, or {\em friction}.

We denote by $-\sigma_{ij}$ the flux of the $i$ component of the
momentum in the $j$ direction.  The tensor $\left[ \sigma \right]$ is
called the {\em viscous stress tensor}.  The $i$ component of the
viscous force per unit volume is then:

\begin{equation}
F_{visc,i} \equiv \frac{\partial \sigma_{ij}}{\partial x_j}.
\label{Fvisc}
\end{equation}

Friction occurs only when different parts of the fluid have different
velocities.  Therefore $\sigma_{ij}$ should depend on the velocity
gradients.  If the velocity varies on a scale large compared to the mean
free path, i.e. to the scale over which molecular transport arises, one
can suppose that $\sigma_{ij}$ depends only on the first derivatives
of the velocity with respect to the coordinates.  Furthermore, we
suppose that the dependence is linear, i.e. we limit ourselves to {\em
Newtonian fluids}.  The most general form of $\sigma_{ij}$ is then:

\begin{equation}
\sigma_{ij} \propto \frac{\partial v_i}{\partial x_j} + A
\frac{\partial v_j}{\partial x_i} + B \delta_{ij} \frac{\partial
v_k}{\partial x_k} ,
\end{equation}

\noindent where $A$ and $B$ are constants to be determined, and
$\delta_{ij}$ is the Kronecker symbol.  If the flow is uniformly
rotating in the $(xy)$--plane for instance, we must have
$\sigma_{xy}=0$.  Since $v_x \propto y$ and $v_y \propto -x$ in that
case, that implies $A=1$.  Therefore the tensor $\left[ \sigma
\right]$ is symmetrical.  We note that:

\begin{displaymath}
Tr \left[ \sigma \right] = \left( 2+3B \right) \nab \cdot {\bf v},
\end{displaymath}

\noindent which shows that $Tr \left[ \sigma \right]$ is a measure of
the volume change of a fluid element.  It is an experimental fact that
the stresses which change the volume of a fluid element give different
viscous forces than the stresses that preserve the volume.  Therefore
we rewrite $\sigma_{ij}$ under the form:

\begin{displaymath}
\sigma_{ij} \propto \left( \frac{\partial v_i}{\partial x_j} +
\frac{\partial v_j}{\partial x_i} - \frac{2}{3} \delta_{ij} \nab \cdot
{\bf v} \right) + b \delta_{ij} \nab \cdot {\bf v},
\end{displaymath}

\noindent where the term in bracket on the right hand side is trace
free, i.e. does not modify the volume of a fluid element.

The {\em shear} and {\em bulk} viscosities, that we denote $\eta$ and
$\zeta$ respectively, are then experimentally defined as:

\begin{equation}
\sigma_{ij} = \eta \left( \frac{\partial v_i}{\partial x_j} +
\frac{\partial v_j}{\partial x_i} - \frac{2}{3} \delta_{ij} \nab \cdot
{\bf v} \right) + \zeta \delta_{ij} \nab \cdot {\bf v}.
\label{sigmaij}
\end{equation}

The bulk viscosity is associated with internal degrees of freedom of
the molecules in the fluid.  It becomes negligible if the
equipartition between these different degrees of freedom is reached
over a timescale shorter than the timescale between two collisions.
Furthermore, for a perfect monoatomic gas it can be shown that
$\zeta=0$ (Huang 1987).  Therefore, from now on we shall ignore
it.

With this expression for $\sigma_{ij}$, we can rewrite the $i$
component of the momentum equation~(\ref{momentum}) under the form:

\begin{equation}
\frac{\partial \left( \rho v_i \right)}{\partial t} +
\frac{\partial}{\partial x_j} \left( \rho v_i v_j + T_{ij} \right) =
F_i ,
\label{momentum2}
\end{equation}

\noindent where

\begin{equation}
T_{ij} = P \delta_{ij} - \sigma_{ij}.
\end{equation}

Equation~(\ref{momentum2}) makes is clear that the flux of the $i$
component of the momentum in the $j$ direction, $\rho v_i v_j +
T_{ij}$, has contributions from both advection and molecular transport
(pressure and viscous forces).

For an incompressible fluid in which $\eta$ is constant, the
corresponding equation of motion, derived from equation~(\ref{motion})
with the above expression of the stress tensor, is called the {\em
Navier--Stokes equation}.

We note that since friction converts mechanical energy into heat, the
rate of change of the kinetic energy, $dE_k/dt$, must be negative if
only viscous forces are acting.  That can be shown from
equation~(\ref{momentum2}) to be equivalent to the condition $\eta>0$
(e.g., Landau \& Lifchitz 1987, \S~II.16).

\subsection{Expression of the shear viscosity}
\label{sec:shear1}

The expression~(\ref{sigmaij}) for the stress tensor need not be
exact.  It has been derived phenomenologically assuming that
$\sigma_{ij}$ depends only on a linear combination of the first
derivatives of the velocity with respect to the coordinates.  However,
the kinetic theory applied to dilute gases leads to the same
expression for $\sigma_{ij}$.  Here we are going to derive an
approximate expression for the shear viscosity $\eta$ by using the
kinetic theory of gases, and we will point out that it arises from
correlations between the particle velocities.

\subsubsection{Kinetic theory of gases}

For simplicity, let us assume that the velocity in cartesian
coordinates is ${\bf v} = v_x(y) {\bf e}_x$, where ${\bf e}_x$ is a
unit vector along the $x$--axis.  The expression~(\ref{sigmaij}) for
the stress tensor then gives:

\begin{equation}
\sigma_{xy}=\sigma_{yx}=\eta \frac{d v_x}{d y}.
\label{sigmaxy}
\end{equation}

This relation had already been proposed by Newton and Hooke, but it is
Maxwell who gave the derivation of the viscosity $\eta$ that we
develop now.

In average, a molecule has a collision with another molecule after it
travels through a distance $\lambda$, which is the mean free path of
the particles.  We suppose that after the collision, the momentum of
the molecule is the same as that of its new environment.

Let us consider the momentum which is transported during the time
$\delta t$ across a surface element $\delta S$ perpendicular to the
$y$--axis and with ordinate $y$.  There are $n u \delta t \delta S /6$
particles crossing $\delta S$ from above during $\delta t$, where $n$
is the number density of particles, $u$ is their random (thermal)
velocity relative to the mean flow and the factor 6 comes about
because there are three possible directions for the particles, each
with two orientations.  Each of these particles travel through
$\lambda$ before it suffers a collision below $\delta S$, which
results in its momentum varying by:

$$ m \left[ v_x(y) - v_x(y+\lambda) \right] \simeq -m \lambda
\frac{dv_x}{dy}, $$

\noindent to first order in $\lambda/L$, where $L$ is the scale of
variation of the velocity.  In other words, each particle carries
below $\delta S$ the {\em excess} of momentum $m \lambda dv_x / dy$.
Here $m$ is the mass of a particle.  On the other hand, each particle
traveling upward carries above $\delta S$ the {\em deficit} of
momentum $-m \lambda dv_x / dy$.  Therefore, the net $x$--component
of the momentum which is carried downward during $\delta t$ by the
particles crossing $\delta S$ is:

\begin{equation}
\delta^2 p_x = 2 \left( \frac{1}{6} n u \delta t \delta S \right)
\left( m \lambda \frac{dv_x}{dt} \right) = \frac{1}{3} n m u
\lambda \frac{dv_x}{dy} \delta S \delta t.
\end{equation}

Let us now consider a box with horizontal faces at $y$ and $y+\delta
y$ and surface area $\delta S$.  The exchange of particles across the
upper face during the time $\delta t$ results in the momentum
$\delta^2 p_x(y+\delta y)$ being {\em added} to the volume, whereas
the exchange across the lower surface results in the momentum
$\delta^2 p_x(y)$ being {\em removed} from the volume.  Therefore, the
time rate of change of the momentum content of the box is:

\begin{equation}
\frac{1}{\delta t} \left[ \delta^2 p_x(y+\delta y) - \delta^2
p_x(y) \right] \simeq \frac{d}{dy} \left( \frac{1}{3} n m u
\lambda \frac{dv_x}{dy} \right) \delta S \delta y 
\end{equation}

\noindent to first order in $\lambda/L$.  This is also $F_{visc,x} \;
\delta S \delta y$, where $F_{visc,x}$ is the viscous force per unit
volume.  Therefore:

\begin{equation}
F_{visc,x} \simeq \frac{d}{dy} \left( \frac{1}{3} n m u
\lambda \frac{dv_x}{dy} \right) .
\end{equation}

Since $F_{visc,x} = d \sigma_{xy}/dy$ and using
equation~(\ref{sigmaxy}), this leads to:

\begin{equation}
\eta \simeq \frac{1}{3} \rho u \lambda ,
\label{eta}
\end{equation}

\noindent where $\rho=nm$ is the mass density of the particles. We also define the {\em kinematic viscosity} $\nu \equiv \eta/\rho$, i.e.

\begin{equation}
\nu \simeq \frac{1}{3} u \lambda.
\label{nu}
\end{equation}

The thermal velocity $u$ is on the order of the sound speed.

We denote by ${\cal R}$ the {\em Reynolds number}, which measures the
relative strength of the inertial and viscous forces.  The inertial
force is $\sim \rho v^2 /L$, where $v$ is the flow velocity, and the
viscous force is $\sim \rho \nu v / L^2$ (from eq.~[\ref{Fvisc}] and
[\ref{sigmaij}]).  Therefore,

\begin{equation}
{\cal R}=L v /\nu.
\label{reynolds}
\end{equation}

\subsubsection{Velocity correlation}

Here again, for simplicity, we consider the case where the flow velocity
is along the $x$--axis and the motion is in the $(xy)$--plane.  The
components of the instantaneous velocity of a particle in the fluid
are $(v_x + u_x, u_y)$, where $u_x$ and $u_y$ are the components of
the random (thermal) velocity relative to the mean flow.  We have
$<u_x>=<u_y>=0$, where the brackets denote a time average.  The flux of
the $x$--component of the momentum along the $y$--direction is:

$$ nm \left( v_x + u_x \right) u_y.$$

Averaged over a large number of particles, or, equivalently, over
time, this gives:

$$ \rho \left< u_x u_y \right> ,$$

\noindent since $<v_x u_y> = v_x <u_y>=0$.  By construction, this
quantity is a component of the stress tensor.  Therefore,

\begin{equation}
\sigma_{xy} = -\rho \left< u_x u_y \right> ,
\label{sigmaxy2}
\end{equation}

\noindent where the minus sign comes about because of the definition
of the viscous stress tensor (see \S~\ref{sec:viscousforces}).  For a
Maxwell--Boltzmann distribution function (or any $xy$ symmetric
function), $<u_x u_y>=0$ and there is no transport.
Equations~(\ref{sigmaxy}) and~(\ref{sigmaxy2}) lead to:

\begin{equation}
\left< u_x u_y \right> = -\nu \frac{d v_x}{d y}.
\end{equation}

Note that in general the time average of the fluctuation velocity may
not be zero.  In an accretion disk for instance, there is a net radial
drift of mass, i.e. the mean value of the radial component of the
fluctuation velocity is finite.  Here, if $<u_y>$ were non zero for
instance, the flux of the $x$--component of the momentum along the
$y$--direction would have the extra contribution $\rho v_x <u_y>$.
This represents the transport of mean momentum by the fluctuations.
Formally however, what we call the stress tensor would still be given
by equation~(\ref{sigmaxy2}), as this is the only contribution from
{\em friction} between particles.

We denote by ${\cal C}$ the correlation coefficient between the
velocities $u_x$ and $u_y$:

$${\cal C} \equiv \frac{\left< u_x u_y \right> }{u^2},$$

\noindent where $u$ is the characteristic velocity of the particles in
the fluid relative to the mean flow.  We see from~(\ref{sigmaxy2})
that ${\cal C}$ gives a measure of the stress tensor.  If $c_s$ is the
sound speed, then $u \sim c_s$.  Therefore

$$ {\cal C} \sim \frac{\nu}{c_s^2} \frac{d v_x}{d y} \sim
\frac{\lambda}{L} \frac{v_x}{c_s} = \frac{\lambda}{L} {\cal M},$$

\noindent where we have used equation~(\ref{nu}) and $d v_x / d y \sim
v_x/L$.  Here ${\cal M}=v_x/c_s$ is the Mach number.  Using
equations~(\ref{nu}) and~(\ref{reynolds}), we can write ${\cal R}=
{\cal M} L/\lambda$, and therefore ${\cal C} = {\cal M}^2/{\cal R}$.

In an accretion disk, ${\cal M} \sim 10$--20 and the Reynolds number
corresponding to the molecular viscosity is larger than $10^{14}$, so
that ${\cal C} \sim 10^{-12} \ll 1$!!

This is a consequence of the small value of the ratio of the mean free
path to the scale of the mean flow ($\sim 10^{-12}$) and it means the
state of the gas is not affected at all by the molecular transport of
angular momentum.  In other words, the flow and the molecular
transport are completely decoupled.

\section{Angular momentum transport by turbulence}

The realization that molecular transport of angular momentum is so
inefficient led the theorists to look for another mechanism of
transport in accretion disks.  Because Reynolds numbers are so high,
it was thought that probably accretion disks would be subject to the
same hydrodynamical nonlinear instabilities that shear flows
experience in laboratory.  The resulting turbulence would then
transport angular momentum efficiently.  Although today much doubt has
been cast on hydrodynamical instabilities in disks, turbulence is
still a strong candidate for transport since it has been shown
relatively recently that a linear magnetohydrodynamical instability
can develop in disks (see below).  Therefore, we turn now to turbulent
transport, and contrast it with molecular transport.  Much of this
section is based on Tennekes \& Lumley (1972), which the reader is
referred to for more details (see also Balbus \& Hawley 1998).

We shall restrict ourselves here to the study of incompressible flows,
as this simplifies the discussion.  For our argument, it is sufficient
to take into account only pressure and viscous forces, but in
principle any other (external or inertial) force could be added.  The
equations describing the fluid are:

\begin{equation}
\frac{\partial {\tilde v}_i}{\partial t} + {\tilde v}_j 
\frac{\partial {\tilde v}_i}{\partial x_j} = \frac{1}{\rho} 
\frac{\partial}{\partial x_j}{\tilde \sigma}_{ij},
\label{motiont}
\end{equation}

and 

\begin{equation}
\frac{\partial {\tilde v}_i}{\partial x_i}=0,
\label{incompt}
\end{equation}

\noindent where the tilde symbol above a variable means we consider
the instantaneous value of the variable at the location $x_i$ and time
$t$.  Here 
$$ {\tilde \sigma}_{ij} = - {\tilde p} \delta_{ij} + \eta {\tilde
s}_{ij}, $$ with $$ {\tilde s}_{ij} = \frac{\partial {\tilde
v}_i}{\partial x_j} + \frac{\partial {\tilde v}_j}{\partial x_i}$$ and
${\tilde p}$ is the pressure.  We suppose $\eta$ is a constant.

We now use the so--called Reynolds decomposition, in which an
instantaneous value is written as the sum of a mean value (denoted by
a capital letter) plus a fluctuation (denoted by a small letter):
$${\tilde v}_i = V_i + v_i .$$ This fluctuation is characteristic of
the turbulence.  This decomposition is meaningful only if the
timescale on which the fluctuations vary and the evolution timescale
of the flow are well separated.  The mean values are then taken over a
timescale large compared to the turbulence timescale but short
compared to that of the flow evolution.  As here we are not interested
in the long term evolution of the flow, we neglect the derivative with
respect to time of the mean values (i.e. we consider a quasi--steady
state).  To simplify the discussion, we suppose that the average of
$v_i$ over time is zero: $<v_i>=0$.

Equation~(\ref{incompt}) averaged over time then leads to $ \partial
V_i / \partial x_i =0,$ i.e. the mean flow is incompressible.
Equation~(\ref{incompt}) thus implies $ \partial v_i / \partial x_i=0,
$ i.e. the fluctuations are also incompressible.

Using $$ {\tilde \sigma}_{ij} = \Sigma_{ij} + \sigma_{ij}$$ with
$\left< \sigma_{ij} \right> = 0$, the equation of motion averaged over
time gives:

\begin{equation}
\frac{\partial}{\partial x_j} V_j V_i + 
\frac{\partial}{\partial x_j} \left< v_j v_i \right> = 
\frac{1}{\rho} \frac{\partial}{\partial x_j} \Sigma_{ij},
\label{momentumt1}
\end{equation}

\noindent where we have used $\partial V_i / \partial t =0$ and the
incompressibility of the mean flow and the fluctuations.  The term
$<v_j v_i >$ represents the averaged transport of the fluctuations of
the momentum by the fluctuations of the velocities.  This is the
turbulent transport.  It is non zero only if the turbulent velocities
in the different directions are correlated.  It is an experimental
fact that this is in general the case for shear turbulence (see
below).  Equation~(\ref{momentumt1}) shows that momentum is
transferred between the fluctuations and the mean flow through the
term $\partial < v_j v_i > / \partial x_j$.

We can rewrite equation~(\ref{momentumt1}) under the form:

\begin{equation}
\frac{\partial}{\partial x_j} {V_j V_i} = \frac{1}{\rho} 
\frac{\partial}{\partial x_j} \left( \Sigma_{ij} + \tau_{ij}
\right),
\label{momentumt2}
\end{equation}

\noindent where $\tau_{ij} = - \rho \left< v_j v_i \right>$ is called
the {\em Reynolds stress tensor}, or turbulent stress.  This is the
contribution from the fluctuations to the averaged total stress tensor
$\Sigma_{ij}+\tau_{ij}$.  Note that if we had not supposed $<v_i>=0$,
we would also have had terms like $V_j <v_i>$, representing transport
(or advection) of mean momentum by the fluctuations.  This is what
Lynden--Bell \& Kalnajs (1972) call {\em lorry--transport}, because it
is a direct "shipment" by the equilibrium flow.  Formally, this term
is not part of what we call the turbulent stress however.

As we have no expression for the components of $\tau_{ij}$ (six of
which are independent), the problem has more unknowns than equations.
This is the well--known closure problem of turbulence.  Since
$\tau_{ij}$ appears in the same way as $\Sigma_{ij}$ in
equation~(\ref{momentumt2}), it is tempting to express $\tau_{ij}$ by
analogy with $\Sigma_{ij}$, which depends on the molecular motion as
we have seen in the previous section.  This is the basis for the {\em
mixing length theory}, in which $\tau_{ij}$ is written exactly like
the tensor deriving from molecular motions using a so--called
turbulent viscosity $\nu_T$.  By analogy with expression~(\ref{nu})
for the molecular viscosity, it is supposed that $\nu_T \sim v_T
\Lambda$, where $v_T$ is a characteristic velocity of the turbulent
eddies and $\Lambda$ is the so--called mixing length, which is the
``mean free path'' of the eddies, i.e. the distance they travel
through before they mix with their environment.

The same analogy is used in accretion disk theory through the $\alpha$
model that we shall describe in details below.

It is important to note that the basis for this analogy is very weak.
For a thorough discussion of the differences between molecular and
turbulent transport, we refer the reader to Tennekes \& Lumley
(1972).  Here we recall the main points. 

There is a fundamental difference between molecular and turbulent
transport, in that turbulence is part of the {\em flow} whereas
molecules are part of the {\em fluid}.  We saw above that molecular
transport is decoupled from the state of the flow.  This is not true
for turbulent transport, which depends completely on the flow for its
existence.  Turbulence can be sustained only if the eddies are able to
tap the energy of the mean flow.  Observations of laboratory shear
flows for instance indicate that the eddies which are most efficient
in tapping the energy of the mean flow are also most efficient in
preserving a good correlation between the different components of the
turbulent velocities, thus giving a large stress tensor.  This is very
different than for the molecules, whose velocities do not
depend on non local properties of the mean flow.

And as a matter of fact, transport of momentum is observed to be a
characteristic of turbulent shear flows.  In the air for instance, the
correlation coefficient between the velocities is on the order of
$10^{-6}$ for the molecules and $0.4$ for turbulent motions.  This is
a characteristic of most shear flows.

In addition, we saw when we derived the expression of the
viscosity~(\ref{eta}) that this requires $\lambda \ll L$.  For a
turbulent flow, that would mean that the scale of the turbulent eddies
is very small compared to the scale over which the characteristics of
the flow, like the mean velocity, vary.  This is not necessarily
satisfied in an accretion disk, where both may be on the order of the
disk semi--thickness.

The conclusion is that shear turbulence is indeed a very efficient way
of transporting momentum, but great care should be taken when
expressing the resulting stress tensor by analogy with molecular
transport.  

In particular, while representing gross transport by an effective
viscosity can often be useful, doing a detailed stability analysis on
a viscous fluid model for turbulent flow is generally not
self--consistent, and can be very misleading (e.g. Hawley, Balbus, \&
Stone 2001).

Note that the above discussion applies to {\em shear} turbulence only,
i.e. flows where the fluctuations get their energy from the mean
velocity gradients.  Transport of momentum may not be nearly as
efficient when the energy source for the fluctuations is a gradient of
temperature or magnetic field for instance.  As a matter of fact,
there are strong indications that the transport of momentum associated
with thermal convection is orders of magnitude weaker than that
associated with shear turbulence (Balbus \& Hawley 1998 and references
therein).

Although turbulence has been considered as a way of transporting
angular momentum in accretion disks for more than fifty years, it is
only relatively recently that an instability able to extract the
energy of the shear and put it in the fluctuations has been
elucidated.  This instability, which requires a magnetic field, is
described in the next section.

\section{Magnetohydrodynamic instabilities}

We consider now the stability of a disk containing a magnetic field.
Balbus \& Hawley (1991, 1998 and references therein) have shown that
in under very general conditions a magnetized disk is subject to a
linear instability which nonlinear development was subsequently found
to result in turbulence (Hawley, Gammie \& Balbus 1995; Brandenburg et
al. 1995).  Following the presentation by Balbus \& Hawley (1998), we
first briefly describe the waves which propagate in a fluid containing
a magnetic field (more details can be found in, e.g., the textbooks by
Jackson~1975, Shu~1992 or Sturrock~1994), and we then go on to
describe the linear instability.

\subsection{Waves propagating in a magnetized fluid}

Here we take into account only pressure and magnetic forces, as these
are the only important forces for our argument.  The governing
equations are the induction equation:

\begin{equation}
\frac{\partial {\bf B}}{\partial t} = \nab \tim \left( {\bf v}
\tim {\bf B} \right),
\label{inductionb}
\end{equation}

\noindent which ensures that $\nab \cdot {\bf B} = 0$, the equation of
continuity:

\begin{equation}
\frac{\partial \rho}{\partial t} + \nab \cdot \left( \rho {\bf v}
\right) =0,
\label{massb}
\end{equation}

\noindent and the equation of motion:

\begin{equation}
\frac{\partial {\bf v}}{\partial t} + \left( {\bf v} \cdot \nab
\right) {\bf v} = - \frac{1}{\rho} \nab P + \frac{1}{\mu_0 \rho}
\left( \nab \tim {\bf B} \right) \tim {\bf B} ,
\label{motionb}
\end{equation}

\noindent where the last term on the right hand side is the Lorentz
force per unit volume in SI units ($\mu_0$ is the permeability of
vacuum).  Note that it can also be written:

\begin{equation}
\frac{1}{\mu_0} \left( \nab \tim {\bf B} \right) \tim {\bf B} = - \nab
\left( \frac{B^2}{2 \mu_0} \right) + \left( \frac{{\bf B}}{\mu_0}
\cdot \nab \right) {\bf B}
\end{equation}

\noindent which shows that the magnetic force if the sum of a magnetic
pressure and a magnetic tension (first term and second term on the
right hand side, respectively).

We first consider a nonrotating fluid and we look for small
perturbations to an equilibrium state characterized by the quantities
$\rho_0$, ${\bf v}_0={\bf 0}$, $P_0$ and ${\bf B}_0$.  We denote the
Eulerian perturbations by a prime, so that $\rho=\rho_0+\rho'$ etc,
and we suppose they vary on a scale much smaller than the scale of
variation of the equilibrium quantities.  The linearized
equations~(\ref{inductionb}), (\ref{massb}) and~(\ref{motionb}) are
then:

\begin{eqnarray}
\frac{\partial {\bf B'}}{\partial t} & = & \nab \tim 
\left( {\bf v'} \tim {\bf B_0} \right) , \label{inductionbp} \\
\frac{\partial \rho'}{\partial t} & + & \rho_0 \nab \cdot {\bf v'} =0, \\
\frac{\partial {\bf v'}}{\partial t} 
& = & - \frac{1}{\rho_0} \nab P' + \frac{1}{\mu_0 \rho_0} \left( \nab
\tim {\bf B'} \right) \tim {\bf B_0}. 
\end{eqnarray}

Since the coefficients of the above equations are constant, we perform
a Fourier transform for the variables ${\bf r}$ and $t$ and look for
solutions under the form exp$\left[ i \left( \omega t - {\bf k} \cdot
{\bf r} \right) \right]$ with $\omega$ and $|{\bf k}|$ real
constants.  The above equations can then be rewritten by replacing
$\partial/\partial t$ by $i\omega$ and $\nab$ by $-i{\bf k}$.

To close the system of equations, we need an equation of state for the
fluid.  We consider adiabatic perturbations (i.e. we suppose there is
no dissipation of heat), so that:

\begin{equation}
\frac{P'}{P} = \gamma \frac{\rho'}{\rho} ,
\label{statebp}
\end{equation}

\noindent where $\gamma$ is the ratio of specific heats ($\gamma=5/3$
for a perfect gas).

The system of equations~(\ref{inductionbp})--(\ref{statebp}) gives a
dispersion relation of order six.  It can be simplified by noting that
the solutions associated with motions perpendicular to the $({\bf k},
{\bf B}_0)$ plane and those associated with motions in this plane can
be decoupled.

The first type of solutions is an {\em Alfven wave}, and its
dispersion relation is:

\begin{equation}
\omega^2 = k^2 v_A^2 \cos^2 \psi ,
\end{equation}

\noindent where $\psi$ is the angle between ${\bf k}$ and ${\bf B}_0$
and $${\bf v}_A = \frac{{\bf B}_0}{\sqrt{\mu_0 \rho_0} }$$ is the
Alfven speed.  These are transverse waves (with no motion in the
$({\bf k}, {\bf B}_0)$ plane) which propagate along the field lines.
They do not involve any compression across the field lines and their
restoring force is the magnetic tension.  These waves are analogous to
the waves which propagate along a stretched string.  Within the factor
$\cos \psi$, we see that the Alfven speed is the square root of the
magnetic tension divided by a mass density, just like the phase speed
in the case of a string.

The second type of solutions is associated with motions only in the
$({\bf k}, {\bf B}_0)$ plane.  Its dispersion relation is:

\begin{equation}
\omega^2 = \frac{k^2}{2} \left\{ \left( v_A^2 + c_s^2 \right)
\pm \left[ \left( v_A^2 + c_s^2 \right)^2 - 4 v_A^2 c_s^2 \cos^2 \psi 
\right]^{1/2} \right\},
\end{equation}

\noindent where $c_s^2=\gamma P_0/ \rho_0$ is the adiabatic sound
speed.  The upper sign, which gives a larger phase speed $\omega/k$,
corresponds to the {\em fast MHD wave}, whereas the lower sign
corresponds to the {\em slow MHD wave}.  The fast mode is also referred
to as {\em magneto--acoustic wave}.  If either $c_s \ll v_A$, $v_A \ll
c_s$ or $\cos \psi \ll 1$, then
$$\omega_+^2 = k^2 \left( v_A^2 + c_s^2 \right)$$ and $$\omega_-^2 =
\frac{k^2 v_A^2 c_s^2 \cos^2 \psi}{v_A^2 + c_s^2},$$ where the
subscripts + and - denote the fast and slow waves, respectively.  For
the fast mode, the magnetic and thermal pressures act in phase.  If
$v_A \gg c_s$, the slow mode is an acoustic wave propagating along the
field lines, whereas if $v_A \ll c_s$ it degenerates into an Alfven
mode in its dispersion properties (the eigenvector is distinct from that
of the Alfven mode however, as the motions are not in the same plane).
For the fast mode it is the opposite.

In the absence of rotation, these modes are stable.  However, Balbus
\& Hawley have shown that when rotation is introduced, the slow mode
can become unstable, and this what we describe now.

\subsection{Linear magnetorotational instability}

We consider the simplest system in which the instability can develop.
This is an axisymmetric disk with a vertical uniform magnetic field.
For the original presentation, which includes the case of a radial
field, see Balbus \& Hawley (1991), and for the stability of a
toroidal field see Balbus \& Hawley (1992b), Fogglizo \& Tagger
(1995), Terquem \& Papaloizou (1996) and Ogilvie \& Pringle (1996).

Since it is the slow mode which is destabilized, one can consider an
incompressible fluid. That amounts to taking the limit $c_s
\rightarrow \infty$ in the above equations, so that the fast mode
disappears (its frequency becomes infinite).

The system of equations describing the fluid is then

\begin{equation}
\frac{\partial {\bf B}}{\partial t} =  \nab \tim \left( {\bf v} \tim
{\bf B} \right), 
\end{equation}

\begin{equation}
\nab \cdot {\bf v}  =  0, 
\end{equation}

\begin{equation}
\frac{\partial {\bf v}}{\partial t} + \left( {\bf v} \cdot \nab \right) 
{\bf v}  =  -\frac{1}{\rho} \nab P + \frac{1}{\mu_0 \rho} \left( \nab 
\tim {\bf B} \right) \tim {\bf B} - \nab \Psi ,
\end{equation}

\noindent where $\Psi$ is the gravitational potential due to the
central star.

We use the cylindrical coordinates $(r, \phi, z)$ and we denote the
unit vectors in the three directions by ${\bf e}_r$, ${\bf e}_{\phi}$
and ${\bf e}_z$.  The components of the vector $\left( {\bf v} \cdot
\nab \right) {\bf v}$ which appears in the equation of motion are:

$$ v_r \frac{\partial v_r}{\partial r} +
\frac{v_{\phi}}{r} \frac{\partial v_r}{\partial \phi} +
v_z \frac{\partial v_r}{\partial z} - \frac{v_{\phi}^2}{r} , $$
$$ v_r \frac{\partial v_{\phi}}{\partial r} +
\frac{v_{\phi}}{r} \frac{\partial v_{\phi}}{\partial \phi} +
v_z \frac{\partial v_{\phi}}{\partial z} + \frac{v_r v_{\phi}}{r} , $$
$$v_r \frac{\partial v_z}{\partial r} +
\frac{v_{\phi}}{r} \frac{\partial v_z}{\partial \phi} +
v_z \frac{\partial v_z}{\partial z}. $$

The equilibrium quantities, that we suppose uniform, are $\rho_0$,
$P_0$, and ${\bf B}_0 = B_0 {\bf e}_z$.  The velocity is ${\bf v}_0 =
r \Omega(r) {\bf e}_{\phi}$.  We now suppose that this equilibrium is
slightly perturbed, so that $\rho=\rho_0+\rho'$ etc, where the prime
denotes the Eulerian perturbations, and we look for solutions
proportional to exp$\left[ i \left( \omega t - k_z z - k_r r \right)
\right]$.  Here we consider axisymmetric perturbations, but more
general solutions can be obtained (see the above mentioned papers).

The linearized system of equations is then:

\begin{eqnarray}
i \omega B'_r & = & - i k_z B v'_r , \\
i \omega B'_{\phi} & = & - i k_z B v'_{\phi} - ik_z r \Omega B'_z +
\left( \frac{d \left( r \Omega \right)}{d r} -i k_r r \Omega \right) 
B'_r , \\
i \omega B'_z & = & - \frac{B}{r} v'_r + ik_r B v'_r, \\
\left( \frac{1}{r} - ik_r \right) v'_r - ikv'_z & = & 0 , \\
i \omega v'_r - 2 \Omega v'_{\phi} & = & \frac{i k_r P'}{\rho} 
- \frac{ik_z BB'_r}{\mu_0 \rho} 
+ \frac{ik_r BB'_z}{\mu_0 \rho}, \\
i \omega v'_{\phi} + 
\left( \frac{d \left( r \Omega \right)}{d r} + \Omega  \right) v'_r
& = & - \frac{ik_z BB'_{\phi}}{\mu_0 \rho} , \\
i \omega v'_z  & = & \frac{ik_z P'}{\rho} ,
\end{eqnarray}

\noindent where, for simplicity, we have dropped the subscript '0' for
the equilibrium quantities.  We now consider large vertical
wavenumbers, such that $|k_z| \gg |k_r|$ and $|k_z| \gg 1/r$. Then,
the above system of equations leads to the following dispersion
relation:

\begin{equation}
\omega^4 - \omega^2 \left( 2k^2v_A^2 + \frac{d \Omega^2
}{d {\rm ln} r} + 4 \Omega^2  \right) + k^2 v_A^2 \left( k^2
v_A^2 + \frac{d \Omega^2}{d {\rm ln} r} \right) = 0 .
\label{dispersionb}
\end{equation}

There is instability if $\omega$ has a negative imaginary part (the
perturbation then grows exponentially with time).
Equation~(\ref{dispersionb}) is a quartic for $\omega^2$ which
solutions are real.  Therefore there is instability if $\omega^2<0$,
which requires:

\begin{equation}
k^2 v_A^2 < - \frac{d \Omega^2}{d {\rm ln} r} .
\label{criterium}
\end{equation}

This criterion has a very simple physical explanation.  It states that
there is an instability when the magnetic tension that acts on a
segment of a field line is smaller than the net tidal force
(i.e. centrifugal force minus gravitational force) acting on it.

For a given equilibrium field $B$, and therefore a given Alfven speed
$v_A$, there will always be a wavenumber $k$ such that this inequality
is satisfied provided the right hand side is positive.  Therefore,
all the disks with

\begin{equation}
\frac{d \Omega^2}{d {\rm ln}r} < 0 
\end{equation}

\noindent are unstable, and this is the criterion for instability.

The heart of the instability resides in the fact that a perturbed fluid
element tends to conserve its angular velocity when a magnetic field
is present.  This is to be contrasted with a non magnetized disk, in
which a perturbed element tends to conserve its specific angular
momentum.  When displaced inward therefore, it has too much angular
momentum for its new location (as the angular momentum increases
outward in an accretion disk), and it moves back to its initial
unperturbed position.  When a magnetic field is present, the magnetic
tension along the field line tends to enforce isorotation of the
elements to which it is connected.  A fluid element displaced inward
has therefore a lower angular velocity than the elements at its new
location, and thus not enough angular momentum for its new position.
As a result it sinks further in.  On the opposite, a fluid element
connected to the same field line and displaced outward will tend to
move further out.  Angular momentum is transferred {\em via} magnetic
torques from the inner fluid element to the outer fluid element.  Note
that the source of free energy for the instability is not in the
magnetic field, but in the disk differential rotation.  The magnetic
field just provides a path to extract the energy.  

From equation~(\ref{dispersionb}), we can write the negative values of
$\omega^2$ as a function of $k^2 v_A^2$, and show that the maximum
growth rate is:

\begin{equation}
\left| \omega_{{\rm max}} \right| = \frac{1}{2}
\left| \frac{d \Omega}{d {\rm ln}r} \right|.
\end{equation}

\noindent   For a Keplerian disk,

\begin{equation}
\left| \omega_{{\rm max}} \right| = \frac{3}{4} \Omega 
\end{equation}

\noindent and is attained for

\begin{equation}
k v_A = \frac{\sqrt{15}}{4} \Omega. 
\end{equation}

This holds even if the field has radial and azimuthal components and
if the perturbed quantities are allowed to vary with $r$ and $\phi$
provided we then replace $k v_A$ by ${\bf k} \cdot {\bf v}_A$.  Note
that the non--axisymmetric case is more subtle though, as in that case
plane waves cannot be sustained, being sheared out by the differential
rotation.  If we write ${\bf k} = k_r {\bf e}_r + (m/r) {\bf e}_{\phi}
+ k_z {\bf e}_z$, with $m$ being the azimuthal wavenumber, then $k_r$
is time dependent and $$k_r(t)=k_0 - mt \frac{d \Omega}{dr}$$
(Goldreich \& Lynden--Bell 1965), which means that a disturbance
always becomes trailing in a disk where the angular velocity decreases
outward.  If $k_r$ is initially large and positive (leading
disturbance), then the mode is stable as indicated
by~(\ref{criterium}).  But as time goes on, $k_r$ decreases, so that
${\bf k}$ enters a region of instability.  As $k_r$ becomes negative
and keeps decreasing though, the mode becomes stable again.  Formally,
the instability is therefore not purely exponential.  The important
question however is whether the mode can be amplified significantly
before its wavelength becomes small enough to be affected by ohmic
resistivity.  This, of course, depends on the magnetic Reynolds
number.

We have seen above that the so--called {\em magnetorotational} or {\em
Balbus--Hawley instability} can develop in any disk in which the
angular velocity decreases outward.  In principle there is no other
condition.  However, it may be that the scale of the modes which are
unstable according to ~(\ref{criterium}) do not fit into the disk,
i.e.  they have a wavelength larger than the disk semithickness $H$.
This is the case if $v_A/\Omega > H$.  Since in a thin disk $\Omega H
\sim c_s$ (see \S~\ref{sec:diskstructure}), the disk will be stable if
$v_A > c_s$.

Another condition which is implicit in the above presentation is, of
course, that the magnetic field be coupled to the fluid.  This may not
be the case everywhere in protostellar disks, which are rather cold
and dense (Gammie 1996; Fromang, Terquem \& Balbus 2001).

Ohmic resistivity can prevent the development of the instability if
the scale on which it acts is comparable to the wavelength of the
unstable modes.  The time it takes for the magnetic field to diffuse
over a scale $\sim 1/k$ due to the effect of an Ohmic resistivity
$\eta_B$ is $\sim 1/(\eta_B k^2)$.  Since in the absence of
resistivity it would grow on a timescale $\sim 1/\Omega$, we see that
Ohmic dissipation prevents the instability if $\eta_B \sim \Omega
/k^2$, i.e. if the magnetic Reynolds number is of order unity.
For more details, the reader is referred to Balbus \& Blaes (1994),
Jin (1996) and Papaloizou \& Terquem (1997).  Note that Fleming, Stone
\& Hawley (2000) have shown that even though the linear instability is
not affected at higher values of the magnetic Reynolds number, the
nonlinear instability cannot be sustained for values as high as $10^4$
in some circumstances, which depend on the field topology for
instance.  However, the issue is not yet settled as other effects like
Hall electromotive forces may resurrect the instability in regions
where Ohmic resistivity acting alone would inhibit it (Wardle 1999,
Balbus \& Terquem 2001).

\subsection{Angular momentum transport in magnetized disks}

Numerical simulations have shown that this instability puts the energy
it extracts from the disk differential rotation into fluctuations
which transport angular momentum outward (see Balbus \& Hawley 1998
and references therein).  As noted above, transport of momentum is not
a property of all types of turbulence.  Furthermore, when it happens,
it is not always against the velocity gradient, i.e. outward in an
accretion disk.  That has to be the case in an isolated dissipative
system, but not if the energy driving the turbulence has an external
source.  We have already mentioned that the transport associated with
thermal convection appears to be very weak compared to that associated
with shear turbulence.  In addition, there are strong indications that
convection driven by external heating transports angular momentum {\em
inward} in accretion disks (Ryu \& Goodman 1992, Cabot 1996, Stone \&
Balbus 1996).  In the case of the magnetorotational instability
though, it can be shown that in the linear regime the transport is
outward (Balbus \& Hawley 1992a).  This still holds in the nonlinear
regime, as expected since the energy source for the turbulence is
internal.  Numerical simulations also show that most of the transport
is done by the (magnetic) Maxwell stress, which dominates over the
(hydrodynamic) Reynolds stress.  Furthermore, magnetic fields are
regenerated through a {\em dynamo} action so that the mechanism can
sustain itself in an isolated system.

It is important to realize that this instability and the resulting
turbulence do not tend to make the disk angular velocity uniform.
This is because the angular velocity is imposed by the gravitational
potential of the object around which the disk rotates.  The turbulence
can smear the shear out only on scales smaller than the disk
semithickness $H$.  To smear the shear out on a scale $l$, the
turbulent velocity $v_t$ has indeed to be on the order of the shear
over $l$, i.e. $l |r d\Omega/dr| \sim l \Omega$, where $\Omega$ is the
angular velocity in the disk.  If $l \sim H$, then $v_t \sim c_s$, as
in a thin disk $c_s \sim H \Omega$ (see \S~\ref{sec:diskstructure}).
Therefore, only supersonic turbulence could modify the shear on scales
larger than $H$.  However, such a turbulence could not be maintained
as the supersonic fluctuations would dissipate into shocks.  The state
toward which the disk evolves as a result of the turbulence is not a
state of uniform rotation but one where all the mass is at the center
and all the angular momentum at infinity (Lynden--Bell \& Pringle
1974).

Long before a mechanism for producing turbulence in accretion disks
had been identified, Shakura \& Sunyaev (1973) proposed a
prescription for modeling turbulent disks.  We now describe this
prescription, and discuss its validity in the context of magnetic
turbulence.  We then develop disk models based on this so--called
$\alpha$ prescription.

\section{$\alpha$ disk models}

\subsection{Evolution of turbulent disks}
\label{sec:evolution}

Here we consider an axisymmetric disk rotating around a central
object.  We suppose the motion is in the plane of the disk, or,
equivalently, we use the vertically averaged equations of mass
conservation and motion.  The velocity is ${\bf v} \equiv (u_r, r
\Omega + u_{\phi})$. The term $r \Omega$ is the circular velocity
around the central mass, and $(u_r, u_{\phi})$ are the components of the
fluctuation velocity.  Note that as the disk accretes, there is a net
radial drift and the mean value of $u_r$ is non zero.  Here, the
equation of mass conservation~(\ref{mass1}) and the azimuthal component
of the equation of motion~(\ref{motion}) are:

\begin{equation}
\frac{\partial \Sigma}{\partial t} + \frac{1}{r}
\frac{\partial}{\partial r} \left( r \Sigma v_r \right) = 0 ,
\label{massd}
\end{equation}

\begin{equation}
\Sigma \left( \frac{\partial v_{\phi}}{\partial t} + v_r
\frac{\partial v_{\phi}}{\partial r} + \frac{v_r v_{\phi}}{r} \right)
= 0 ,
\label{motiond}
\end{equation}

\noindent where $\Sigma$ is the surface mass density, i.e. the mass
density integrated over the disk thickness.  Here, according to the
discussion in \S~\ref{sec:shear1}, we have neglected the viscous force
arising from molecular viscosity.  Multiplying
equation~(\ref{motiond}) by $r$ and using~(\ref{massd}), we obtain the
angular momentum equation:

\begin{equation}
\frac{\partial}{\partial t} \left( r \Sigma \left( r \Omega + u_{\phi}
\right) \right) + \frac{1}{r} \frac{\partial}{\partial r} \left( r^2
\Sigma \left( r \Omega + u_{\phi} \right) u_r \right) = 0 .
\label{momentumd1}
\end{equation}

As pointed out by Balbus \& Papaloizou (1999), to get a diffusion
equation describing the disk evolution we need to smooth out the
fluctuations over radius.  To do so, we average the above equation
over a scale large compared to that of the fluctuations, but small
compared to the characteristic scale of the flow (i.e. $r$).
Equation~(\ref{momentumd1}) then yields:

\begin{equation}
\frac{\partial}{\partial t} \left( \Sigma r^2 \Omega \right) +
\frac{1}{r} \frac{\partial}{\partial r} \left( \Sigma r^3 \Omega <u_r>
+ \Sigma r^2 < u_r u_{\phi} > \right) = 0 ,
\label{momentumd}
\end{equation}

\noindent where the brackets denote the radial average and we have
neglected $<u_{\phi}>$ compared to $r \Omega$ in the time derivative.
This is justified because $|<u_{\phi}>| \ll r \Omega$ and the
systematic, evolutionary time derivative of $<u_{\phi}>$ is limited.
In the radial divergence term however, both $<u_r>$ and $<u_r
u_{\phi}>$ are second order, and therefore we retain all the terms.
This equation is the same as that describing a viscous flow with ${\bf
v} \equiv (u_r, r\Omega)$ and a stress tensor $\sigma_{r \phi} \equiv
- \Sigma < u_r u_{\phi} >$ as given by equation~(\ref{sigmaxy2}).

There are two contributions to the flux of angular momentum: the term
$\Sigma r^2 \Omega <u_r>$ is the mean angular momentum advected
through the disk by the velocity fluctuations (because of the
accretion of mass), whereas the term $\Sigma r < u_r u_{\phi} >$
represents the angular momentum fluctuations transported by the
velocity fluctuations.

Using equation~(\ref{massd}) averaged over radius, we can rewrite
equation~(\ref{momentumd}) under the form:

$$ r \Sigma <u_r> = - \left[ \frac{d}{dr} \left( r^2 \Omega \right)
\right]^{-1} \frac{\partial}{\partial r} \left( \Sigma r^2 < u_r
u_{\phi} > \right).$$

\noindent Using equation~(\ref{massd}) again to eliminate $<u_r>$,
this leads to the diffusion equation:

\begin{equation}
\frac{\partial \Sigma}{\partial t} = \frac{1}{r}
\frac{\partial}{\partial r} \left(
\left[ \frac{d}{dr} \left( r^2 \Omega \right)
\right]^{-1} \frac{\partial}{\partial r} \left( \Sigma r^2 < u_r
u_{\phi} > \right) \right) .
\label{diffusion}
\end{equation}

In a steady state, equation~(\ref{massd}) gives $r \Sigma <u_r> = {\rm
constant}$.  Then the accretion rate

\begin{equation}
\dot{M} \equiv -2 \pi r \Sigma <u_r>
\label{dotM}
\end{equation}

\noindent is constant through the disk.  Integration of the angular
momentum equation~(\ref{momentumd}) then yields:

\begin{equation}
\Sigma < u_r u_{\phi} > = \frac{\dot{M}}{2 \pi} \Omega \left[ 1 -
\left( \frac{R_i}{r} \right)^{1/2} \right] ,
\label{dotM2}
\end{equation}

where $R_i$ is the disk inner boundary.  We have assumed here that the
turbulent stress $< u_r u_{\phi}>$ vanishes at the disk inner edge
(i.e. there is no torque at the boundary) and that the disk is
Keplerian, so that $\Omega \propto r^{-3/2}$.  The above equation
shows that for the mass to be accreted inward (i.e. $<u_r>$ negative),
the flux of angular momentum due to the fluctuations has to be
positive, i.e.  the fluctuations have to transport angular momentum
outward.

\subsection{The $\alpha$ prescription}

We pointed out that the angular momentum equation~(\ref{momentumd}) is
analogous to that describing a viscous flow with ${\bf v} \equiv (u_r,
r\Omega)$ and a stress tensor $\sigma_{r \phi} \equiv - \Sigma < u_r
u_{\phi} >$.  Therefore, it is tempting to push the analogy further
and express the turbulent stress $- \Sigma < u_r u_{\phi} >$ as if it
derived from an enhanced 'turbulent viscosity' $\nu$, defined such
that (equation~[\ref{sigmaij}]):

\begin{equation}
- \Sigma < u_r u_{\phi} > \equiv \Sigma \nu r \frac{d\Omega}{dr} .
\end{equation}

In a Keplerian disk, this gives:

\begin{equation} 
< u_r u_{\phi} > \sim  \nu \Omega. 
\label{uruphi1}
\end{equation}

The equations presented in the previous section have been derived by
Lynden--Bell \& Pringle (1974; see also Pringle 1981) assuming a
viscous flow with this expression of the stress tensor.

The prescription proposed by Shakura \& Sunyaev (1973) consists in
writing $\nu = v_t H$, where $H$ is the disk thickness, assumed to be
the maximum scale of the turbulent cells, and $v_t$ is the turbulent
velocity.  They further define $$\alpha \equiv \frac{v_t}{c_s},$$
where $c_s$ is the sound velocity.  Note that $\alpha<1$, otherwise
the fluctuations would dissipate into shocks in such a way as to
restore $v_t<c_s$.  Equation~(\ref{uruphi1}) can then be rewritten
under the form:

\begin{equation}
< u_r u_{\phi} > \sim \alpha c_s^2.
\label{uruphi}
\end{equation}

\noindent Here we have used the fact that in a thin disk $H \sim c_s /
\Omega$ (see \S~\ref{sec:diskstructure}).

So far we have focussed on non magnetized disks.  In these, there are
strong indications that $<u_r u_{\phi}>$ is either zero or negative.
Magnetism is needed to correlate the velocities.  The above discussion
does apply to magnetized disks provided we replace $< u_r u_{\phi} >$
by $< u_r u_{\phi} - u_{Ar} u_{A \phi}>$, where $(u_{Ar}, u_{A \phi})$
are the components of the fluctuations of the Alfven velocity (Shakura
\& Sunyaev 1973, Balbus \& Hawley 1998).  The extra term represents
the Maxwell stress.

The validity of the $\alpha$ prescription in the context of magnetic
turbulence was discussed by Balbus \& Papaloizou (1999).  They first
pointed out that, as long as $<u_r u_{\phi}>$ (or $< u_r u_{\phi} -
u_{Ar} u_{A \phi}>$) is positive, the disk dynamics is the same as if
it were evolving under the action of a viscosity.  In that case
indeed, the diffusion coefficient in equation~(\ref{diffusion}) is
positive.  We can then always define an $\alpha$ parameter according
to ~(\ref{uruphi}), although $\alpha$ may not be constant through the
disk.

Note however that this implicitly assumes it is possible to average
the equations over radius in the way described in
\S~\ref{sec:evolution}.  If the scale of the fluctuations and the
characteristic disk scale are not well separated, such an average
cannot be performed.  Since the maximum scale of the fluctuations is
of order the disk thickness $H$, this procedure requires that there is
a scale large compared to $H$ and small compared to $r$.  This
condition may be only marginally fulfilled in protostellar disks, in
which $H$ may be up to 0.1--$0.2r$.

Balbus \& Papaloizou (1999) further noted that for the $\alpha$
prescription to apply, the disk had to behave viscously not only in
its {\em dynamics} but also in its {\em energetics}.  The key point
here is that a viscous disk dissipates {\em locally} the energy it
extracts from the shear, whether in a steady state or not.  This may
not be the case in a turbulent disk where, if the turbulent cascade is
not efficient, part of the energy may be advected with the flow.  As
we have not addressed the energetics of viscous disks above, we will
not go into the details of the discussion here.  These can be found in
Balbus \& Papaloizou (1999), who showed that in disks subject to MHD
turbulence the energy extracted from the shear is indeed dissipated
locally (through the turbulent cascade) whether the disk is evolving
or not.  Note that this is in general not the case when the turbulence
is due to self--gravitating instabilities.  In that case indeed, part
of the energy is transported by waves through the disk.

Most theoretical protostellar disk models have relied on the $\alpha$
parametrization.  We have commented that MHD turbulence does lend
itself to this prescription in thin disks.  However, because MHD
instabilities develop only in an adequately ionized fluid, they may
not operate everywhere in protostellar disks (Gammie 1996).  Therefore
it is likely that the parameter $\alpha$ is not constant through these
disks.  It may even be that only parts of these disks can be described
using this $\alpha$ prescription.  However, we may still learn about
disks from these models in the same way as we learned about stars from
simple polytropic models. Therefore, we now describe these models.

\subsection{Vertical structure}

In a thin disk, the radial gradients of temperature and pressure are
much smaller than the vertical gradients.  The vertical and radial
structures therefore decouple, and can be calculated separately.  Here
we present the calculation of the vertical structure, i.e. we
determine pressure, temperature and density as a function of height
$z$.  Once this is done, given an initial surface density
distribution, the radial structure can be computed using the diffusion
equation~(\ref{diffusion}).  The section below is based on Papaloizou
\& Terquem (1999), which the reader is referred to for more details.

\subsubsection{Basic equations}
\label{sec:diskstructure}

The equations describing the disk vertical structure are the equation
of vertical hydrostatic equilibrium:

\begin{equation}
\frac{1}{\rho} \frac{\partial P}{\partial z} = - \frac{G M_{\ast} z}
{\left( r^2 + z^2 \right) ^{3/2}} ,
\label{dPdz}
\end{equation}

\noindent and the energy equation, which states that the rate of
energy removal by radiation is locally balanced by the rate of energy
production by viscous dissipation:

\begin{equation}
\frac{\partial {\cal F}}{\partial z} = \frac{9}{4} \rho \nu \Omega^2 ,
\label{dFdz1}
\end{equation}

\noindent where ${\cal F}$ is the radiative flux of energy through a
surface of constant $z$  which is given by:

\begin{equation}
{\cal F} = \frac{- 16 \sigma T^3}{3 \kappa \rho}
\frac{\partial T}{\partial z} .
\label{dTdz}
\end{equation}

\noindent Here $G$ is the gravitational constant, $M_{\ast}$ is the central
mass, $P$ is the pressure, $T$ is the temperature, $\kappa$ is the
opacity, which in general depends on both $\rho$ and $T$, and $\sigma$
is the Stefan--Boltzmann constant.

\noindent To close the system of equations, we relate $P$, $\rho$ and
$T$ through the equation of state of an ideal gas: 

\begin{equation}
P = \frac{\rho k T}{\mu m_H} ,
\label{state}
\end{equation}

\noindent where $k$ is the Boltzmann constant, $\mu$ is the mean
molecular weight and $m_H$ is the mass of the hydrogen atom. If we
limit our calculations to temperatures lower than 4,000~K (typical of
protostellar disks), then, at the densities of interest, hydrogen is
in molecular form.  Since the main component of protostellar disks is
hydrogen, we then have $\mu=2$.  We note that for the temperatures we
consider transport of energy by convection can be neglected.

We denote the isothermal sound speed by $c_s$ ($c_s^2=P/\rho$).  We
adopt the $\alpha$--parametrization of Shakura \& Sunyaev (1973), so
that the kinematic viscosity is written $\nu=\alpha c_s^2/\Omega$
(equations~[\ref{uruphi1}] and~[\ref{uruphi}]). In general, $\alpha$
is a function of both $r$ and $z$.  With this formalism,
equation~(\ref{dFdz1}) becomes:

\begin{equation}
\frac{\partial {\cal F}}{\partial z} = \frac{9}{4} \alpha \Omega P .
\label{dFdz}
\end{equation}

Note that, in a thin disk, $z \ll r$.  Then, if we set $|\partial P
/\partial z| \sim P/H$ and $z \sim H$ in equation~(\ref{dPdz}), we get
$c_s \sim \Omega H$, where $\Omega$ is the Keplerian velocity
($\Omega^2 = GM_{\ast}/r^3$) and we have used $c_s^2=P/\rho$.  The thin
disk approximation then requires $c_s \ll r \Omega$, i.e. the angular
velocity to be highly supersonic.

\subsubsection{Boundary conditions} 

We have to solve three first order ordinary differential equations for
the three variables ${\cal F}$, $P$ (or equivalently $\rho$), and $T$
as a function of $z$ at a given radius $r.$ Accordingly, we need three
boundary conditions at each $r$. We shall denote with a subscript $s$
values at the disk surface.

The first boundary condition is obtained by integrating
equation~(\ref{dFdz1}) over $z$ between $-H$ and $H$. Since by
symmetry ${\cal F}(z=0)=0$, this gives:

\begin{equation}
{\cal F}_s = \frac{3}{8 \pi} \dot{M}_{st} \Omega^2 ,
\label{Fs}
\end{equation}

\noindent where we have defined $\dot{M}_{st} \equiv 3 \pi
\langle{\nu}\rangle \Sigma$, with $<\nu>$ being the vertically
averaged viscosity.  If the disk were in a steady state,
$\dot{M}_{st}$ would not vary with $r$ and would be the constant
accretion rate through the disk (eq.~[\ref{dotM2}] and~[\ref{uruphi1}]
with $R_i \ll r$). In general however, this quantity does depend on
$r$.

Another boundary condition is obtained by integrating
equation~(\ref{dPdz}) over $z$ between $H$ and infinity.  A detailed
derivation of this condition is presented in the Appendix~A of
Papaloizou \& Terquem (1999).  Here we just give the result:

\begin{equation}
P_s = \frac{\Omega^2 H \tau_{ab}}{\kappa_s} ,
\label{Ps}
\end{equation}

\noindent where $\tau_{ab}$ is the optical depth above the disk.  This
condition is familiar in stellar structure, where $\Omega^2 H$ would
be replaced by the acceleration of gravity at the stellar surface.
Since we have defined the disk surface such that the atmosphere above
the disk is isothermal, we have to take $\tau_{ab} \ll 1$.  Providing
this is satisfied, the results do not depend on the value of
$\tau_{ab}$ we choose.

A third and final boundary condition is given by the expression of the
surface temperature (see Appendix~A of Papaloizou \& Terquem 1999 for
a detailed derivation of this expression):

\begin{equation}
2 \sigma \left( T_s^4 - T_b^4 \right) - \frac{9 \alpha k T_s \Omega}{8
\mu m_H \kappa_s} - \frac{3}{8 \pi} \dot{M}_{st} \Omega^2 = 0 .
\label{Ts}
\end{equation}

\noindent Here the disk is assumed immersed in a medium with
background temperature $T_b$.  The surface opacity $\kappa_s$ in
general depends on both $T_s$ and $\rho_s$ and we have used
$c_s^2=kT/(\mu m_H)$.  The boundary condition~(\ref{Ts}) is the same
as that used by Levermore \& Pomraning (1981) in the Eddington
approximation (their eq.~[56] with $\gamma=1/2$).  In the simple case
when $T_b=0$ and the surface dissipation term involving $\alpha$ is
set to zero, with $\dot{M}_{st}$ being retained, it simply relates the
disk surface temperature to the emergent radiation flux.

\subsubsection{ Disk models}

At a given radius $r$ and for a given value of the parameters
$\dot{M}_{st}$ and $\alpha$ (which uniquely determine the disk model),
the disk vertical structure is obtained by solving
equations~(\ref{dPdz}), (\ref{dTdz}) and~(\ref{dFdz}) with the
boundary conditions~(\ref{Fs}), (\ref{Ps}) and~(\ref{Ts}).  

Numerical results are published, e.g., by Papaloizou \& Terquem (1999)
(see also Lin \& Papaloizou 1980, 1985; Bell \& Lin 1994; Bell et
al. 1997), in which the opacity is taken from Bell \& Lin (1994). This
has contributions from dust grains, molecules, atoms and ions.  It is
written in the form $\kappa=\kappa_i \rho^a T^b$ where $\kappa_i$, $a$
and $b$ vary with temperature.  For a specified $\dot{M}_{st},$ they
calculate the value of $H,$ the vertical height of the disk surface,
iteratively.  Starting from an estimated value of $H,$ after
satisfying the surface boundary conditions, the equations are
integrated down to the mid--plane $z=0.$ The condition that ${\cal
F}=0$ at $z=0$ is not in general satisfied.  An iteration procedure is
then used to adjust value of $H$ until ${\cal F}=0$ at $z=0$ to a
specified accuracy.

\section*{Acknowledgments}

It is a pleasure to thank Steven Balbus and John Papaloizou for
sharing with me their knowledge and thoughts about accretion disk
theory, and for their comments on an early draft of this paper which
led to much improvements.

\end{document}